# Semicircular-aperture illumination scanning transmission electron microscopy

*Akira Yasuhara[1*], Fumio Hosokawa[2*], Sadayuki Asaoka[3], Teppei Akiyama[3], Tomokazu Iyoda[4], Chikako Nakayama[1], and Takumi Sannomiya[5*]*

[1] JEOL Ltd., 3-1-2 Musashino, Akishima, Tokyo, 196-8558, Japan.

[2] FH electron optics, 1-5-18-805, Kita, Kunitachi,Tokyo, 186-0001, Japan.

[3] Faculty of Materials Science and Engineering, Kyoto Institute of Technology, Matsugaskaki, Sakyo-ku, Kyoto 606-8585, Japan.

[4] Faculty of Science and Engineering, Doshisha University, Kyotanabe, Kyoto 610-0394, Japan.

[5] Department of Materials Science and Engineering, School of Materials and Chemical Technologies, Institute of Science Tokyo, 4259 Nagatsuta, Midoriku, Yokohama, 226-8503, Japan.

**Corresponding Authors**

\* Akira Yasuhara (Email: ayasuhar@jeol.co.jp)

\* Fumio Hosokawa (Email: sim_em_image@jcom.zaq.ne.jp)

\* Takumi Sannomiya (Email: sannomiya.t.aa@m.titech.ac.jp)

KEYWORDS

semicircular aperture, contrast transfer function, scanning transmission electron microscopy, phase contrast

**ABSTRACT**

Scanning transmission electron microscopy (STEM) provides high-resolution visualization of atomic structures as well as various functional imaging modes utilizing phase contrast. In this study we introduce a semicircular aperture in STEM bright field imaging, which gives a phase contrast transfer function that becomes complex and includes both lower and higher spatial frequency contrast transfer. This approach offers significant advantages over conventional phase plate methods, having no charge accumulation, degradation, or unwanted background noise, which are all problematic in the phase plate material. Also compared to the differential phase contrast or ptychography equipment, this semicircular aperture is far less costly. We apply this approach to visualization of polymer, biological and magnetic samples.

**INTRODUCTION**

Scanning transmission electron microscopy (STEM) and its related imaging techniques have significantly advanced visualization of material structures down to atomic scales as well as material functions.[1] Among the STEM techniques, high-angle annular dark field (HAADF) imaging stands out for the straightforward interpretation of its image contrast, which is based on the convolution of the specimen potential and the point spread function (PSF) of the electron beam, where contrast is heavily influenced by atomic number differences. Dark field imaging can also be performed at low-angle, which detects scattered intensity due to multiple scattering interactions with the real potential, while HAADF imaging involves convolution with an imaginary potential, integrating the loss electrons due to thermal diffuse scattering (attenuated by the Debye-Waller factor) across the detection area.[2-4] STEM bright field (BF) imaging, which linearly detects potentials for thin samples, allows for easier identification of light elements in structures compared to HAADF. However, the interpretation of BF images is complex due to the intricate changes in image formation dependent on defocus. This complexity is identical to the linear imaging components of conventional transmission electron microscopy (TEM), which will be discussed in detail in the theoretical section of this article.

For the phase imaging techniques, contrast retrieval using phase plates has been successful especially in visualization of unstained biological samples for TEM.[5-8] Although similar benefits are anticipated for STEM, challenges such as charge issues, aging, or handling difficulties still seem to limit reports of their use.[9-12] Differential phase contrast (DPC) imaging, although costly because of additional detector sets, provides significant contrast enhancement and has been applied to various imaging including electric potentials as well as magnetic fields.[13-15] More advanced techniques employing more detectors (although even more costly) have allowed imaging of highly-beam sensitive materials by optimizing the contrast transfer[16] or by ptychography.[17, 18]

In this study, we explore an alternative low-cost phase imaging technique by the use of semicircular-aperture illumination (SCI) in STEM imaging. While corresponding Foucault TEM has a long history, the STEM version seems left unexplored.[6, 7, 19] Theoretical considerations reveal that the phase contrast transfer function (PCTF) becomes a complex function, even when only considering even-order aberrations (such as defocus and spherical aberration). The SCI-STEM-BF image comprises a standard BF image with sine type response to the potential and a derivative phase image obtained through Hilbert-transforming the cosine type phase-plate image across the semicircular aperture. This results in an interesting power spectrum without Thon rings at any defocus condition, which is confirmed experimentally. We applied this SCI-STEM approach to image polymer and biological samples for the linear contrast imaging as well as to visualize magnetic domains for the scattering intensity contrast imaging. Since the electron beam is simply blocked by a thick structure like a conventional aperture edge, this approach experiences no degradation or charge accumulation, which is the main concern in the phase plate imaging with thin films. For thin-film phase plates scattering by the material potential of the thin film introduces certain noise components, whereas this SCI-STEM approach includes no such image deterioration.

**THEORY**

*Phase contrast transfer function for STEM*

According to the reciprocity of TEM and STEM, PCTF for BF imaging should be identical for both methods, which we first describe through generic description. The wavefunction on the aperture plane is :

$$\phi_c(k) = A_p(k) \exp\left\{\frac{2\pi i}{\lambda}\chi(k)\right\} \tag{1}$$

For a circular aperture of radius $R$, the aperture function is defined as $A_p(k) = 1$ for $|k| \leq R$ and $A_p(k) = 0$ for $|k| > R$. $\chi(k)$ is the wave aberration. The resultant electron beam spot $B(r,t)$ formed on the sample plane at the beam scan position $t$ becomes

$$B(r,t) = \text{FT}_{k \to r}[\,\phi_c(k)\exp\{2\pi i(-kt)\}\,] \tag{2}$$

Here, the coordinate $k$ is converted to $r$ in real space by Fourier transform FT. Right after the sample with projected potential $V_p(r)$ and interaction coefficient, the wavefunction is

$$\begin{aligned}
\Phi(r,t) &= B(r,t)\exp\{-i\sigma V_p(r)\} \\
&= \text{FT}_{k \to r}[\,\phi_c(k)\exp\{2\pi i(-kt)\}\,] \cdot [\cos\{\sigma V_p(r)\} - i\sin\{\sigma V_p(r)\}] \\
&\sim \text{FT}_{k \to r}[\,\phi_c(k)\exp\{2\pi i(-kt)\}\,] \cdot [1 - i\sigma V_p(r)] \quad .
\end{aligned} \tag{3}$$

The approximation of the last line in Eq.(3) holds for sufficiently small $\sigma V_p(r)$, such as weak phase objects. On the detector plane, the wavefunction is expressed using convolution operation * as

$$\phi_b(k,t) = \mathrm{FT}_{r \to k}[\,\Phi(r,t)\,] = \{\phi_c(-k)\exp(2\pi i k t)\} * \mathrm{FT}_{r \to k}[1 - i\sigma V_p(r)]$$

$$= \phi_c(-k)\exp(2\pi i k t) - \{\phi_c(-k)\exp(2\pi i k t)\} * i\sigma v_p(k)$$

$$= \phi_c(-k)\exp(2\pi i k t) - \int \left[\phi_c(-k+\xi)\exp(2\pi i k t)\, i\sigma v_p(\xi)\exp(-2\pi i \xi t)\right] d\xi$$

$$= \exp(2\pi i k t)\{\phi_c(-k) - \mathrm{FT}^{-1}_{\xi \to t}[\phi_c(-k+\xi)i\sigma v_p(\xi)]\}. \tag{4}$$

$v_p(k)$ is the sample function in *k*-space. When a point detector is placed at position $k_d$, the detection intensity at beam position *t* is

$$|\phi_b(k_d,t)|^2 = |A_p(-k_d)|^2$$

$$- \phi_c(-k_d) \cdot \mathrm{FT}^{-1}_{\xi \to t}\left[\,\overline{\phi_c(-k_d-\xi)i\sigma v_p(-\xi)}\,\right]$$

$$- \overline{\phi_c(-k_d)} \cdot \mathrm{FT}^{-1}_{\xi \to t}[\phi_c(-k_d+\xi)i\sigma v_p(\xi)]$$

$$+ \mathrm{FT}^{-1}_{\xi \to t}[\phi_c(-k_d+\xi)i\sigma v_p(\xi)] \cdot \mathrm{FT}^{-1}_{\xi \to t}\left[\,\overline{\phi_c(-k_d-\xi)i\sigma v_p(-\xi)}\,\right]. \tag{5}$$

Here we used the relation for Fourier transform $\overline{\mathrm{FT}[F(u)]} = \mathrm{FT}[\overline{F(-u)}]$. The first term corresponds to the direct transmitted wave, which is constant, and the fourth term to the interference between the scattered waves. The fourth term can be neglected for a weak object. The second and third terms correspond to the linear term to the potential, which we consider in the following discussions. This linear intensity *I* in the frequency space *u* on the formed image is obtained by Fourier transforming the linear terms from *t* to *u* space:

$$I(k_d, u) = \phi_c(-k_d) \cdot \overline{\phi_c(-k_d - u)} i\sigma v_p(u)$$

$$- \overline{\phi_c(-k_d)} \cdot \phi_c(-k_d + u) i\sigma v_p(u)$$

$$= i\sigma v_p(u)\left[A_p(-k_d)A_p(-k_d - u)\exp\left\{\tfrac{2\pi i}{\lambda}(\chi(-k_d) - \chi(-k_d - u))\right\}\right.$$

$$- A_p(-k_d)A_p(-k_d + u)\exp\left\{\frac{2\pi i}{\lambda}(-\chi(-k_d) + \chi(-k_d + u))\right\}\Big]. \quad (6)$$

The relation $\overline{v_p(-\xi)} = v_p(\xi)$ for the real potential is used. Eq (6) is the general formalism of the STEM BF image for the linear phase contrast, which includes detector position ($k_d$), aperture ($A_p$), and aberration ($\chi$). With a point detector placed on the axis ($k_d = 0$) and without offset phase shift e.g. by phase plate ($\chi(0) = 0$), the BF image intensity is:

$$I(u) = i\sigma v_p(u)\big[A_p(-u)\exp\{-i\gamma(-u)\} - A_p(u)\exp\{i\gamma(u)\}\big] \quad (7)$$

We introduced the phase shift $\gamma(u) = \frac{2\pi}{\lambda}\chi(u)$ by wave aberration. All these expressions of the wavefunctions are schematically summarized in Figure 1.

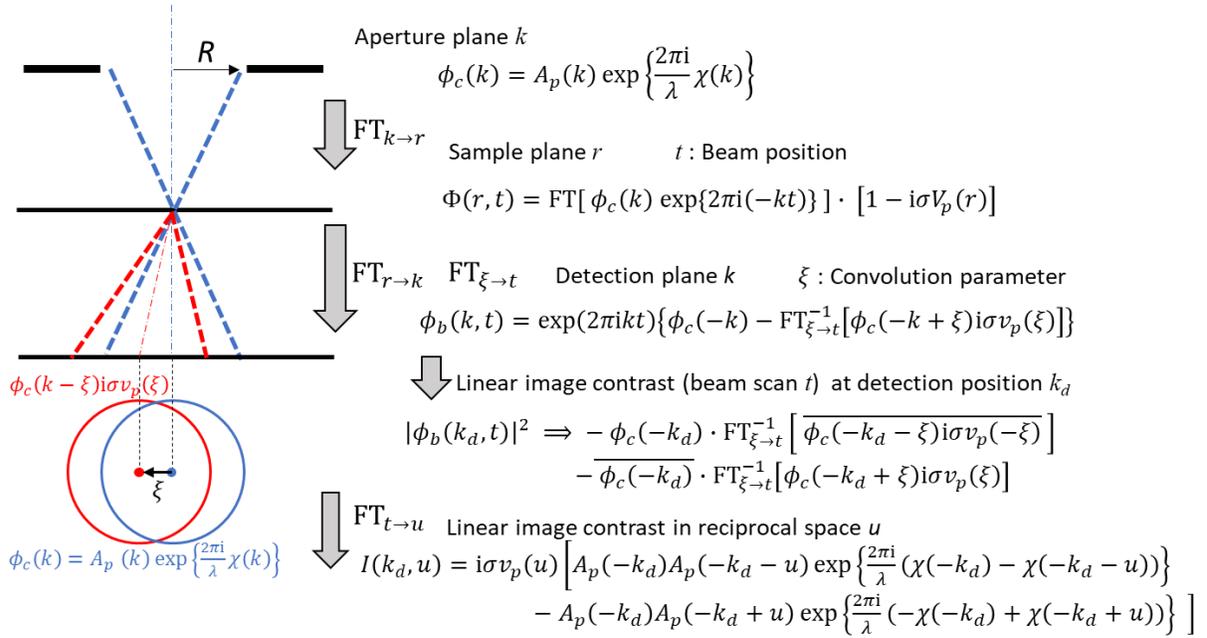

Fig. 1 Schematic diagram of the wavefunction expressed on each optical plane.

*Circular aperture*

With a sufficiently large circular aperture compared to the detector size on the projected detector plane, the aperture function can be considered as $A_p(u) = 1$. The linear contrast of Eq (7) for a circular aperture becomes

$$I(u) = i\sigma v_p(u)[\exp\{-i\gamma(-u))\} - \exp\{i\gamma(u))\}]$$

$$= 2\sigma v_p(u)\left[\sin\left\{\frac{\gamma(u) + \gamma(-u)}{2}\right\}\exp\left\{i\frac{\gamma(u) - \gamma(-u)}{2}\right\}\right]. \quad (8)$$

Eq(8) represents the phase contrast transfer function for conventional STEM BF, which corresponds to TEM BF. When only even aberration functions, such as defocus, two-fold astigmatism, or spherical aberration, are included (i.e. $\gamma(u) = \gamma(-u)$ ), Eq.(7) is reduced to $I(u) = 2\sigma v_p(u)\sin\{\gamma(u)\}$.

*Semicircular aperture*

A semicircular aperture used in SCI-STEM can be defined using the *x*-component $u_x$ in the frequency space $u$ as $A_p(u) = 1$ for $u_x \geq 0$, and $A_p(u) = 0$ for $u_x < 0$. The phase contrast transfer function for semicircular aperture BF image is:

$$\text{for } u_x \geq 0: \quad I(u) = \sigma v_p(u)\exp\left[i\left\{\gamma(u) - \frac{\pi}{2}\right\}\right] \quad (9a)$$

$$= \sigma v_p(u)\sin\left\{\frac{\gamma(u) + \gamma(-u)}{2}\right\}\exp\left\{i\frac{\gamma(u) - \gamma(-u)}{2}\right\}$$

$$-\mathrm{i}\sigma v_p(u)\cos\left\{\frac{\gamma(u)+\gamma(-u)}{2}\right\}\exp\left\{\mathrm{i}\frac{\gamma(u)-\gamma(-u)}{2}\right\} \quad (9\mathrm{b})$$

$$\text{for } u_x < 0: \quad I(u) = \sigma v_p(u)\exp\left[\mathrm{i}\left\{-\gamma(-u)+\frac{\pi}{2}\right\}\right] \quad (10\mathrm{a})$$

$$= \sigma v_p(u)\sin\left\{\frac{\gamma(u)+\gamma(-u)}{2}\right\}\exp\left\{\mathrm{i}\frac{\gamma(u)-\gamma(-u)}{2}\right\}$$

$$+ \mathrm{i}\sigma v_p(u)\cos\left\{\frac{\gamma(u)+\gamma(-u)}{2}\right\}\exp\left\{\mathrm{i}\frac{\gamma(u)-\gamma(-u)}{2}\right\} \quad (10\mathrm{b})$$

The equations (9a) and (10a) express the transfer of the amplitude and phase while (9b) and (10b) show the even and odd aberrations explicitly. These expressions guarantee that the real part of the transfer function is an even function and that the imaginary part is an odd function for any aberrations, which is consistent with Friedel's law. In the case when only even aberrations (defocus, two-fold astigmatism, spherical aberration, etc) are present, these expressions reduce to:

$$\text{for } u_x \geq 0: \quad I(u) = \sigma v_p(u)\sin\{\gamma(u)\} - \mathrm{i}\sigma v_p(u)\cos\{\gamma(u)\}$$

$$\text{for } u_x < 0: \quad I(u) = \sigma v_p(u)\sin\{\gamma(u)\} + \mathrm{i}\sigma v_p(u)\cos\{\gamma(u)\}. \quad (11)$$

From Eq(11), one understands that the contrast transfer function consists of a sum of the conventional BF component (first term) and the component by Hilbert transforming a π/2 phase plate image (Zernike image) along the *x* direction (second term).[20, 21] Thus, the SCI-STEM imaging is capable of retrieving low-frequency phase information akin to the phase-plate imaging.

*Spatial and temporal coherence*

A finite-size circular detector $D(k_d)$ with a radius $Q$ placed on the axis can be defined as $D(k_d) = 1/S$ for $|k_d| \leq Q$ and $D(k_d) = 0$ for $|k_d| > Q$, which degrades the spatial coherence similarly to the source size effect in TEM. The signal is normalized by detector area $S$. Then the signal within such a finite-size detector $I_{coh}$ is obtained by integrating the general intensity expression Eq (6) over the detector area.

$$I_{coh}(u) = \int I(k_d, u) D(k_d) \, dk_d$$

$$= i\sigma v_p(u) \int \left[ A_p(-k_d) A_p(-k_d - u) \exp\left\{\frac{2\pi i}{\lambda}(\chi(-k_d) - \chi(-k_d - u))\right\} \right.$$

$$\left. - A_p(-k_d) A_p(-k_d + u) \exp\left\{\frac{2\pi i}{\lambda}(-\chi(-k_d) + \chi(-k_d + u))\right\} \right] D(k_d) dk_d \quad (12)$$

In the condition that the edge of the aperture images not going across the detector and that $k_d$ is relatively small leading to the aberration being approximated as $\chi(k_d + u) \sim \chi(u) + k_d \nabla \chi(u)$, this can be reduced into an expression with a more explicit damping term,

$$I_{coh}(u) = i\sigma v_p(u) \, A_p(-u) \exp\left\{-\frac{2\pi i}{\lambda}\chi(-u)\right\} \text{FT}_{k_d \to G(-u)}[D(k_d)]$$

$$- i\sigma v_p(u) \, A_p(u) \exp\left\{\frac{2\pi i}{\lambda}\chi(u)\right\} \text{FT}^{-1}_{k_d \to G(u)}[D(k_d)] \, . \quad (13)$$

The transfer function is modulated by the Fourier-transformed detector in the aberration space, which corresponds to the signal damping due to the spatial coherence and the aberration. For a circular detector, this damping can be expressed by first order Bessel function of the first kind $J_1$ as,[22]

$$D_{\mp p}(u) = \text{FT}^{\pm 1}_{k_d \to G(\mp u)}[D(k_d)] = \frac{J_1[2\pi Q |G(\mp u)|]}{\pi Q |G(\mp u)|} \ . \tag{14}$$

Here, we introduced a geometrical aberration $G(u) = \nabla \chi(u)/\lambda$. For the temporal coherence, the energy spread of the source is responsible for the signal damping. Using chromatic aberration $C_c$ and acceleration voltage $U$, defocus change $\Delta_0$ by energy spread $\Delta U$ in the voltage unit is expressed as $\Delta_0 = C_c(\Delta U/U)(1 + 2\gamma U)/(1 + \gamma U)$. Here, $\gamma$ is the relativity correction factor of the value of $9.7847 \times 10^{-7}$ V$^{-1}$. The envelope function $E_d$ then becomes

$$E_d(u) = \exp(-\pi^2 \Delta_0^2 \frac{\lambda^2 u^4}{4}) \ . \tag{15}$$

The phase contrast transfer function considering these damping factors is shown in Figure 2.

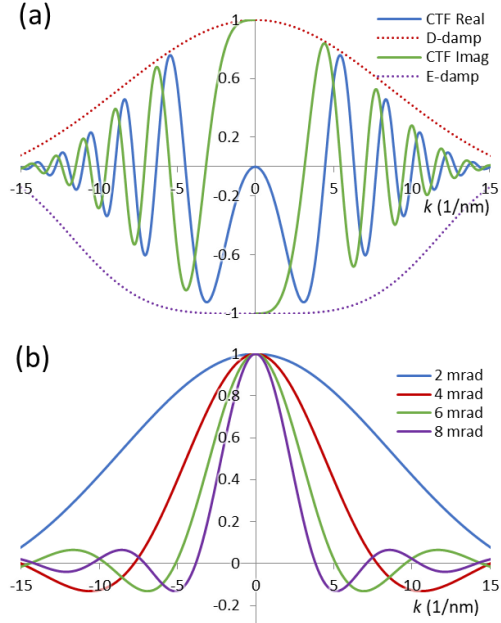

Figure 2. (a) Phase contrast transfer function for a semicircular aperture with damping. The following parameters are used. Acceleration voltage: 200kV, defocus = -20 nm, Cs = 3 μm, $\Delta_0$ = 1.575 nm, $Q$ = 2 mrad. (b) Bessel damping functions with different detector radii $Q$ of 2-8 mrad.

**EXPERIMENTAL METHODS**

*Semicircular aperture*

The semicircular aperture with the radius of the circle of 20 µm was fabricated using a focused ion beam (FIB) method on a 10 µm Au film. Two semicircular apertures with 90° rotation to each other were prepared. The fabricated apertures are installed in the condenser lens aperture module in a JEM-ARM200F (JEOL Ltd., Japan) instrument with an aberration corrector for STEM (Figure 3). The convergence angle on STEM observation was controlled by changing the illumination lens setting of STEM. For comparison, conventional circular apertures with different diameters are also available in this aperture module.

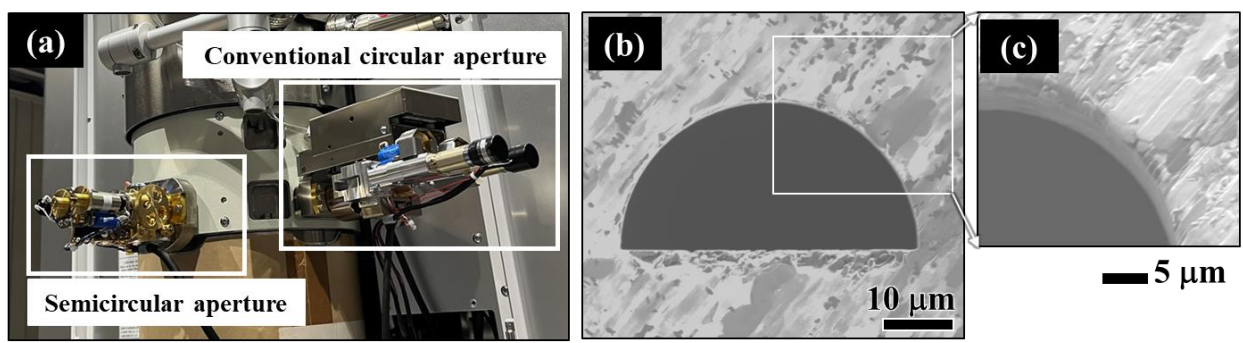

Figure 3. Condenser lens (CL) aperture module in the JEM-ARM200F instrument. (a) Appearance of the CL aperture module. Two sets of the aperture module are installed. Conventional circular apertures are installed in one module, and semicircular apertures are set in the other one. (b, c) Secondary ion microscope images of the top and tilt views of the semicircular aperture fabricated using a FIB method.

*STEM observation*

STEM observations were conducted under two conditions using the JEM-ARM200F instrument: a standard high magnification mode with the objective lens turned on and a low magnification mode with the objective lens off. By these two modes, the convergence semi angles (based on a circular aperture) can range from 28 mrad down to 150 μrad. The residual aberrations are minimized by the aberration corrector although some residual two-fold astigmatism may remain. The detection angle for STEM-BF can be controlled by changing the setting of image forming lenses and using dedicated circular shape apertures with several diameters, which are placed right above the BF-STEM detector. In addition, we also utilized a two-dimensional pixelized detector (4DCanvas, JEOL Ltd.) to investigate the influence of the detector setting. The observation was all performed under 200 kV acceleration.

*STEM simulation*

Elbis software was used for numerical STEM image simulation, which can better reproduce the experimental results while the analytical formulation in the previous section gives intuitive interpretation of the transfer functions. The Elbis simulation is based on the multi-slice method considering multiple scattering from the phase lattice of the electrostatic potentials and can produce both TEM and STEM imaging considering various axial and chromatic aberrations. High-load images, such as amorphous STEM images, can be efficiently simulated by boosting the calculation speed with GPU-based parallel computing.[2]

**RESULTS AND DISCUSSION**

*Phase contrast transfer function*

  PCTF is evaluated by imaging an amorphous sample which includes all the spatial frequencies. Figure 4 shows the STEM-BF images of an amorphous Ge thin film using conventional circular and semicircular apertures for illumination in STEM. A small detector, indicated by red circles in Fig. 4a and h, is set at the center of the illumination aperture. Left two columns show the results using a conventional circular aperture and right two columns those using a semicircular aperture. The aperture images on the aperture plane are shown in the top row (Fig. 4a and h). The second raw (Fig. 4b, e, i and l) shows the original BF images of the amorphous Ge film in two different defocus conditions. The corresponding Fourier transformation (FT) patterns in the third row (Fig. 4c, f, j and m) visualize the PCTF. For the circular aperture (Fig. 4c and f), a typical sinus type modulation is observed, especially with the node and anti-node rings appearing in the defocused condition (Fig. 4f). This nicely corresponds to the expression of Eq. (8) for the circular aperture. This is also reproduced by the FT pattern from the numerically simulated amorphous images, as shown in Fig. 4d and g.

  In contrast, the FT patterns of the semicircular aperture images (Fig.4j and m) show characteristic azimuthal angle dependence, as indicated by orange and blue arrows, which is due to the effect of the finite size aperture; Along the blue arrow direction, which corresponds to the straight chord line of the semicircle, the aperture configuration with respect to the finite size detector is actually same as the circular one since both positive and negative sides are open. Indeed, the profiles of the FT patterns in this blue arrow direction in Fig. 4j and m is similar to those of the conventional ones (Fig. 4c and f). On the other hand, in the orange arrow direction where semi-circular shape effectively works, the FT patterns of the semicircular aperture show a unique PCTF with no oscillating feature even in the defocused condition. This is in accordance with the theoretical expression of Eq (9a) and (10a), where the intensity does not modulate due to the addition of the cosine-type phase contrast component as shown in Eq. (11). These features are also reproduced in the simulated images in Fig. 4d, g, k and n. The aberration damping features due to defocus is visible in Fig. 4f, m, g and n, experimentally representing the characteristics of Eq (13) and (14). Especially, the oscillating feature of the Bessel function becomes visible in the semicircular aperture along the yellow arrows in Fig. 4n, which cannot be recognized in conventional STEM-BF with a circular aperture because of the sinus oscillation of PCTF.

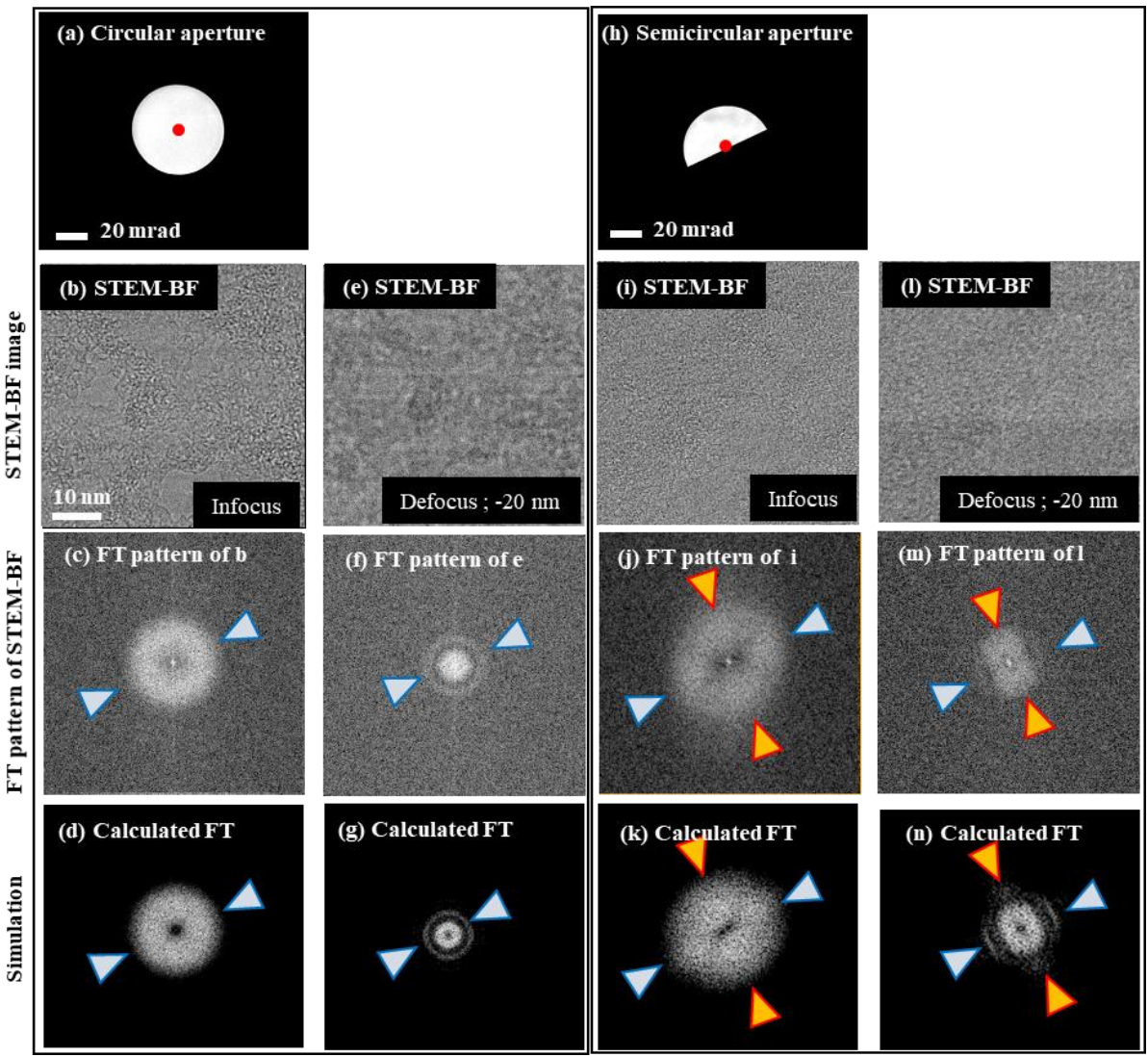

Figure 4. Comparison of PCTFs of conventional and SCI BF images using an amorphous Ge sample. The images of left two columns display the results of a conventional circular aperture and right two columns the results of a semicircular aperture. Images of panels b, e, i and l show the original STEM-BF images with different focuses. Panels c, f, j and m show the corresponding FT patterns. Panels d, g, k and n are FT patterns of simulated STEM images of amorphous Ge using the corresponding settings to the experiment; sample thickness: 3 nm, acceleration voltage: 200 kV, CL aperture radius: 25 mrad on axis, Cs: 1.0μm. The defocus values are (d) -5 nm, (g) -20nm, (k) -5 nm, and (n) -20nm.

*Imaging of organic specimen: block copolymer*

Since the SCI-STEM approach is capable of low frequency phase imaging because of inclusion of the Hilbert phase-plate contrast, we apply this to image an unstained organic polymer material where the low frequency phase contrast imaging becomes critical. We prepared a block-copolymer thin membrane with microphase-segregated nanostructures by depositing the polymers on the water surface and transferred to a TEM grid.[23] (details are described in Supplementary Materials) Figure 5 shows the conventional and SCI STEM-BF images. For SCI imaging two different orientations of the aperture are shown. Because of the beam-sensitive nature of the sample, we could not image the same sample position in different conditions. To maximize the low frequency contrast, the convergence semi-angle was set to be 1 mrad and detection semi-angle 30 μrad. We chose relatively inhomogeneous areas, where the array grains are smaller to involve low-frequency information. Thanks to the low illumination angle, the hexagonal array patterns of the block copolymer is visible even in the conventional STEM image in Fig. 5a. The contrast is much clearer in the SCI images in Fig. 5b and c. The periodicity of the cylindrical structure of the block copolymer is well represented as the ring pattern in FT pattern. It is noticed that the direction of the differential phase contrast of the Hilbert component is rotated by different semicircular aperture directions in Fig. 5b and c, namely differentiation in 4-10 o'clock direction for Fig. 5b and 1-7 o'clock direction in Fig. 5b., which corresponds to the straight chord line of the semicircle. In the inset FT patterns, the lower frequency signals, especially inside the periodicity ring, are enhanced in the SCI approach (panels b and c) in the direction of this differentiation. As discussed above, the contrast transfer is same as the conventional method in the straight chord line direction (perpendicular to the differentiation direction).

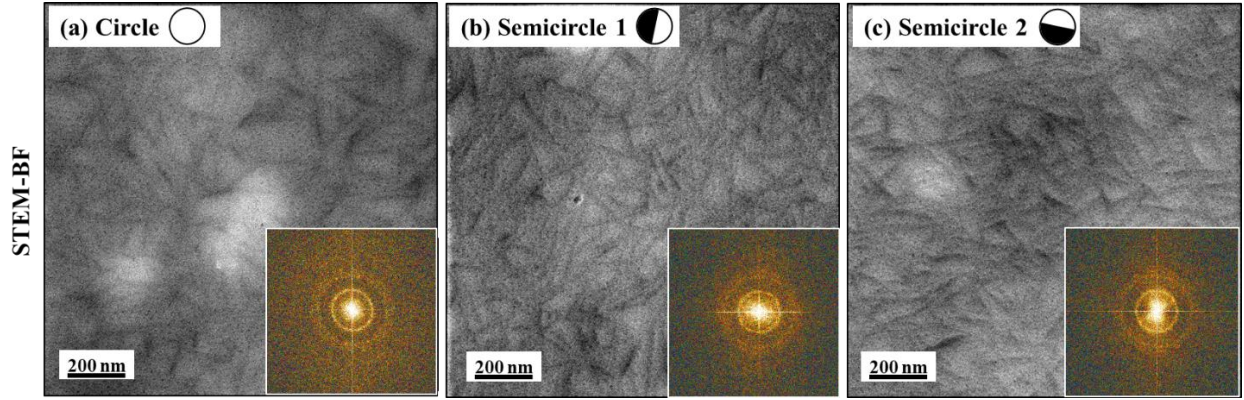

Figure 5. Conventional and SCI STEM-BF images of a block copolymer sample. The FT patterns of each images are inserted at the bottom right. (a) Image with a conventional circular aperture. (b, e) Images with semicircular apertures with different orientations. The configuration of the semicircular aperture is schematically illustrated on top of each image.

*Imaging of biological specimen: mouse liver*

We test the SCI approach also on an unstained biological specimen. Figure 6 shows the observation results of a liver of mouse chemically fixed by a glutaraldehyde and $OsO_4$. An ultrathin section for TEM observation was prepared by the ultra-microtome method after embedding the sample in epoxy resin. For the STEM-BF and SCI observation, the apertures with a convergence semi-angle of ~ 200 µrad are used, as shown in the insets of Fig. 6a and c, where the detector size and position are indicated by red circles. The images in Fig. 6 are taken at the lumen of sinusoid. While liver endothelial structures are visible in both STEM-BF and SCI images, microvilli on the hepatocyte are more clearly imaged in SCI images in Fig. 6c and d, for instance, at the arrow position in Fig. 6d. Notably, the SCI images show enhanced contrasts at the edge of the microvilli around the space of Disse (Fig. 6d) towards the perpendicular direction of the chord line of the aperture, clearly indicating the Hilbert component as described in Eq. 11. A comparison of line profiles of BF and SCI-STEM images are found in the Supplementary Material.

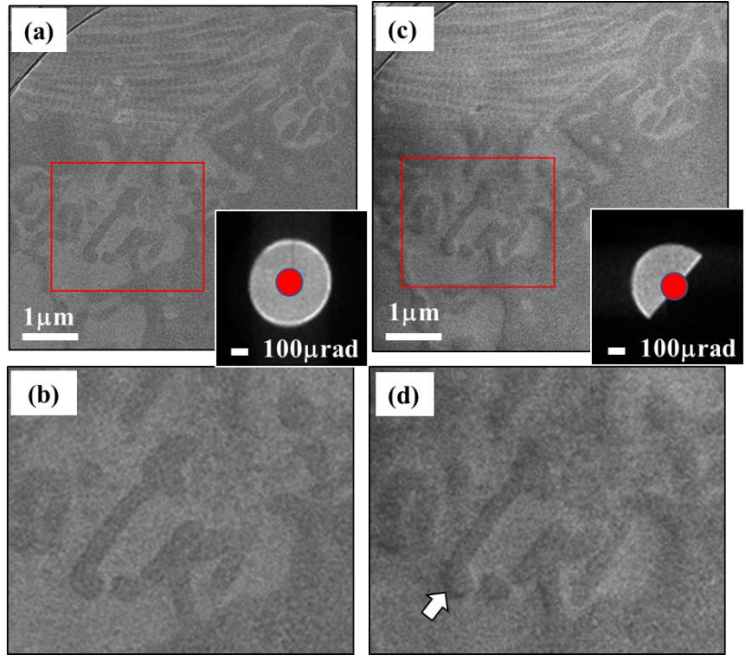

Figure 6. Conventional (a, b) and SCI STEM-BF (c, d) images of a biological specimen. The insets of panels a and c show the aperture images and the detector positions indicated by the red circle. Panels b and d are magnified images of the boxed regions in the images a and b, respectively.

*Magnetic domain observation*

Since the differential contrast of the Hilbert component is available in SCI STEM-BF imaging, we also try visualizing magnetic domains, expecting similar imaging as differential phase contrast (DPC) imaging by segmented detector. However, it is noted that the direct "reference" signal is not present for magnetic domain imaging since all the electron beam is deflected, the contrast formation in the magnetic domain imaging is more the scattering intensity, corresponding to the last term in Eq. (5), than linear interference terms.

We used a permalloy thin film for the domain observation. The result is shown in Figure.7. The objective lens is turned off for this observation to avoid magnetizing the sample. The conventional STEM BF image shown in Fig. 7a shows a uniform contrast on the film. The black frame is the supporting Cu grid. In contrast, SCI STEM-BF imaging with the semicircular aperture allows visualizing the magnetic domains with clear dark and bright contrasts as shown in Fig. 7 b and c. The dark and bright contrasts correspond to opposite directions of the magnetization vector. By rotating the differential direction, i.e. rotating the semicircular aperture by 90°, the domain contrasts of image b and c display different domain features, showing the 90° orthogonal components of magnetization vector on the sample plane. Slight linear background shading may be included due to the imperfect correction of the beam shift by the scan for such low magnification imaging.

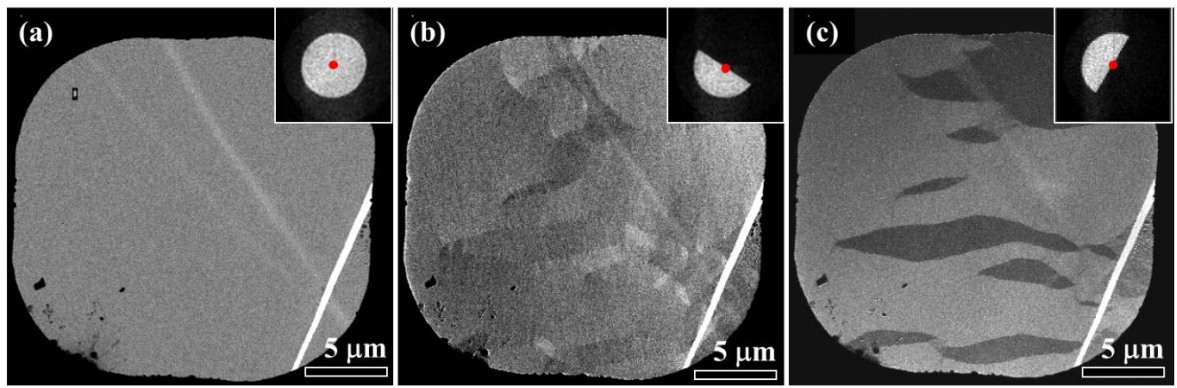

Figure.7 Conventional (a) and SCI STEM-BF (b, c) images of a permalloy thin film for magnetic domain observation. The CL aperture images are shown in the inset of each STEM images (a, b, and c). The red circles at the bright field discs indicate the detector. The convergence semi-angle is 150 μrad.

## CONCLUSIONS

We have introduced the STEM-BF imaging using a semicircular aperture. The image formation for this SCI-STEM imaging is theoretically derived. The obtained PCTF shows that SCI STEM-BF image consists of conventional BF contrast with sinus type signal modulation and Hilbert-phase-plate type component, showing the potential of the low frequency phase contrast retrieval useful for organic or biological imaging. We experimentally demonstrated this imaging by introducing a semicircular aperture in STEM instrument confirmed that the PCTF indeed involves cosine type component resulting in no oscillative PCTF features even when defocus is introduced. We further applied this approach to visualize an organic polymer sample of block copolymer, which nicely showed high-contrast periodic patterns with representative differential contrast, corroborating the usefulness of this approach. The biological imaging using a mouse liver specimen also confirmed the higher contrast imaging capability compared to conventional method. We also successfully visualized magnetic domains using this SCI STEM approach, of which vector component can be separately extracted by semicircular aperture orientations.

The proposed approach is simple in terms of installation since just installing another (or two more for different differentiation directions) suffices and no other additional mechanical or electrical component is needed. Thus, this SCI STEM is an inexpensive and handy alternative to DPC-STEM that require dedicated detector installation, or to phase-plate imaging where degradation or charging of the phase plate material becomes critical in contrast interpretation as well as in reproducibility.

## ACKNOWLEDGMENT

This work is financially supported by JSPS Kakenhi (22H05033, 24H00400) and JST FOREST (JPMJFR213J).

## Author Contributions

The manuscript was written through contributions of all authors. All authors have given approval to the final version of the manuscript.

# Supplementary Material for

# Semicircular-aperture illumination scanning transmission electron microscopy

*Akira Yasuhara[1*], Fumio Hosokawa[2*], Sadayuki Asaoka[3], Teppei Akiyama[3], Tomokazu Iyoda[4], Chikako Nakayama[1], and Takumi Sannomiya[5*]*

[1] JEOL Ltd., 3-1-2 Musashino, Akishima, Tokyo, 196-8558, Japan.

[2] FH electron optics, 1-5-18-805, Kita, Kunitachi,Tokyo, 186-0001, Japan.

[3] Faculty of Materials Science and Engineering, Kyoto Institute of Technology, Matsugaskaki, Sakyo-ku, Kyoto 606-8585, Japan.

[4] Faculty of Science and Engineering, Doshisha University, Kyotanabe, Kyoto 610-0394, Japan.

[5] Department of Materials Science and Engineering, School of Materials and Chemical Technologies, Institute of Science Tokyo, 4259 Nagatsuta, Midoriku, Yokohama, 226-8503, Japan.

*Corresponding Authors*
* Akira Yasuhara (Email: ayasuhar@jeol.co.jp)
* Fumio Hosokawa (Email: sim_em_image@jcom.zaq.ne.jp)
* Takumi Sannomiya (Email: sannomiya.t.aa@m.titech.ac.jp)

**Definition of Fourier transform used in the manuscript**

The Fourier transform (FT) in this article is defined as follows.

$$F(k) = \text{FT}[f(r)] = \int f(r) \exp(2\pi i k r)\, dr \tag{S1}$$

$$f(r) = \text{FT}^{-1}[F(k)] = \int F(k) \exp(-2\pi i k r) \tag{S2}$$

The forward transform (eq. S1) indicates changing the space from $r$ to $k$ with the wave propagation from the source to the screen, whereas the inverse transform (eq. S2) corresponds to the back-propagating trace of the wave, i.e. from the screen to the source.

**Block copolymer sample preparation**

The monomer 11-(4-((*E*)-4-butylstyryl)phenoxy)undecyl methacrylate (MA(Stb)) was purchased from TCI and utilized without further purification. The macroinitiator α-methoxy-poly(ethylene oxide)-ω-(2-bromo-2-methylpropionate) (pEO-BMP) was synthesized through the esterification of poly(ethylene oxide) monomethyl ether ($M_N$ = 4830,NOF) with 2-bromo-2-methylpropionyl bromide (TCI) following a previously reported procedure.[1] $^1$H NMR (D2O) δ 1.92 (s, 6H), 3.34 (s, 3H), 3.54 (m, 2H), 3.59 (m, 2H), 3.60-3.71 (m, (4n – 10)H), 3.78 (m, 4H), 4.34 (m, 2H); $M_N$ = 8760; $M_W/M_N$ = 1.03. The amphiphilic diblock copolymer pEO-b-pMA(Stb) was synthesized via atom-transfer radical polymerization of MA(Stb), initiated by pEO-BMP in the presence of a 1,1,4,7,10,10-hexamethyltriethylenetetramine copper(I) complex catalyst in anisole.1 $M_N$ = 43500; $M_W/M_N$ = 1.16. The degree of polymerization of the pMA(Stb) segment was determined to be 71, calculated from the ratio of the peak areas in the $^1$H NMR spectrum and the MN of the pEO segment. For the TEM sample preparation, the block copolymer thin film was fabricated by dispersing 3.5 wt% toluene solutions of the copolymers onto a water surface, subsequently transferring the thin film onto an elastic carbon-coated copper TEM grid, and subjecting it to thermal annealing at 190°C for 6 hours in a vacuum oven.

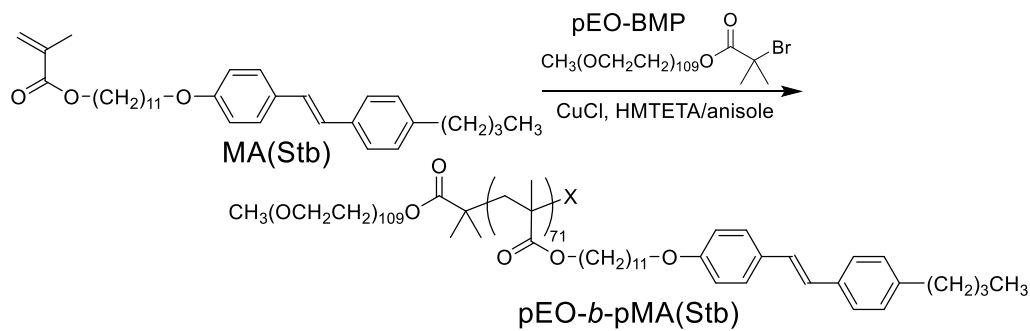

Figure S1. Synthesis reaction of the used block copolymer.

**Simulated images and power spectra**

Figures S2-S5 show simulated STEM-BF images of amorphous Ge in different aperture and detector configurations. The following are the simulation parameters: amorphous Ge sample thickness: 3nm, acceleration voltage: 200kV, CL aperture radius: 25mrad, Cs: 1.0 μm, detector position: on axis with different radii $Q$. With the semicircular aperture (Figs. S3, S5), the "differentiation" direction is horizontal (direction of 3-9 o'clock) in the image, i.e. the chord of the semicircle is oriented along the direction of 12-6 o'clock.

For the semicircular aperture results in Figure S3, the power spectra do not change by defocus change. This supports the analytical prediction of Eqs (9-10) with non-oscillating amplitudes. With this detector size, the coherence damping by the aberration is not significant, which is also clear for the circular aperture with the same detector size (Fig. S2). In contrast, with a larger detector size, the power spectra are significantly influenced by the defocus, as shown in Figs. S5.

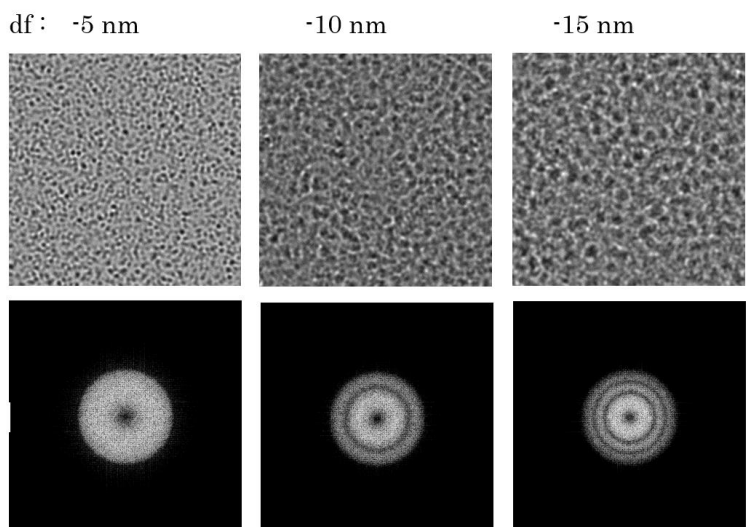

Figure S2. (upper row) STEM-BF images of an amorphous Ge thin film using a circular aperture. The detector semi-angle is $Q = 2$ mrad. (lower row) FT power spectra of the corresponding images.

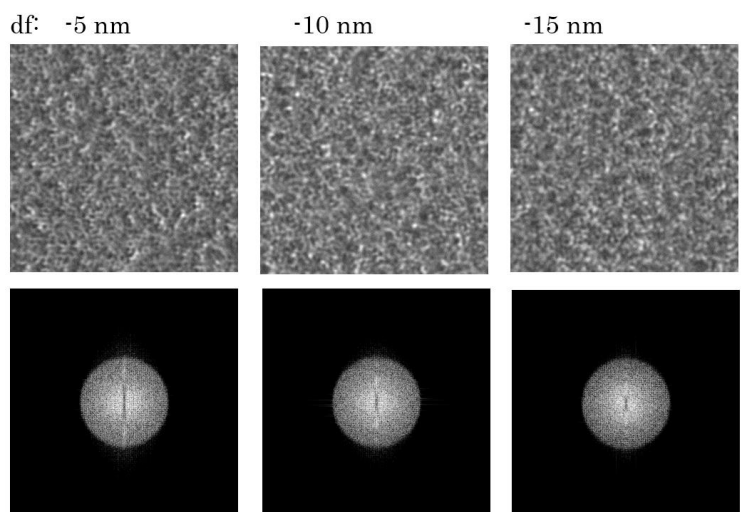

Figure S3. (upper row) STEM-BF images of an amorphous Ge thin film using a semicircular aperture. The detector semi-angle is $Q = 2$ mrad. (lower row) FT power spectra of the corresponding images.

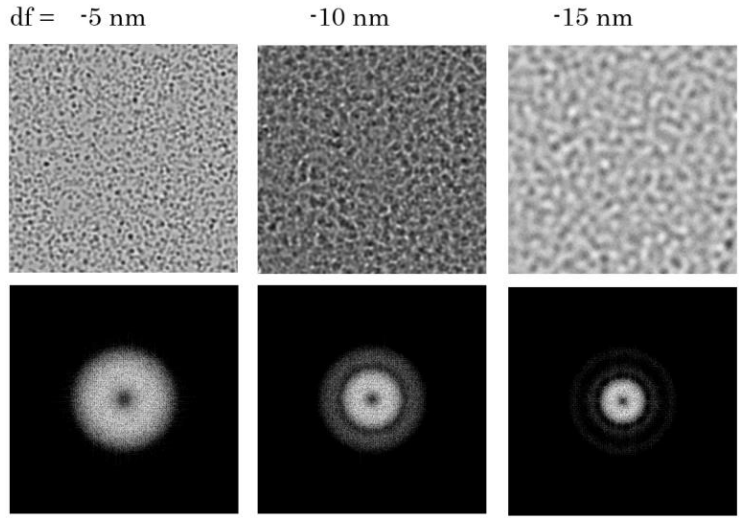

Figure S4. (upper row) STEM-BF images of an amorphous Ge thin film using a circular aperture. The detector semi-angle is $Q = 8$ mrad. (lower row) FT power spectra of the corresponding images.

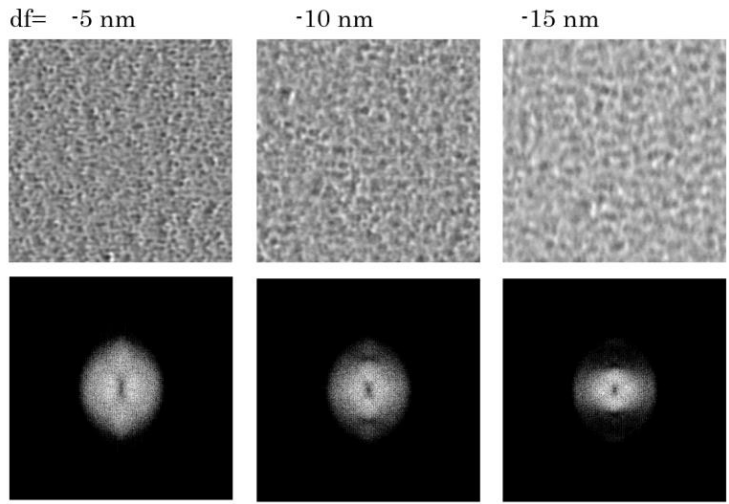

Figure S5. (upper row) STEM-BF images of an amorphous Ge thin film using a semicircular aperture. The detector semi-angle is $Q = 8$ mrad. (lower row) FT power spectra of the corresponding images.

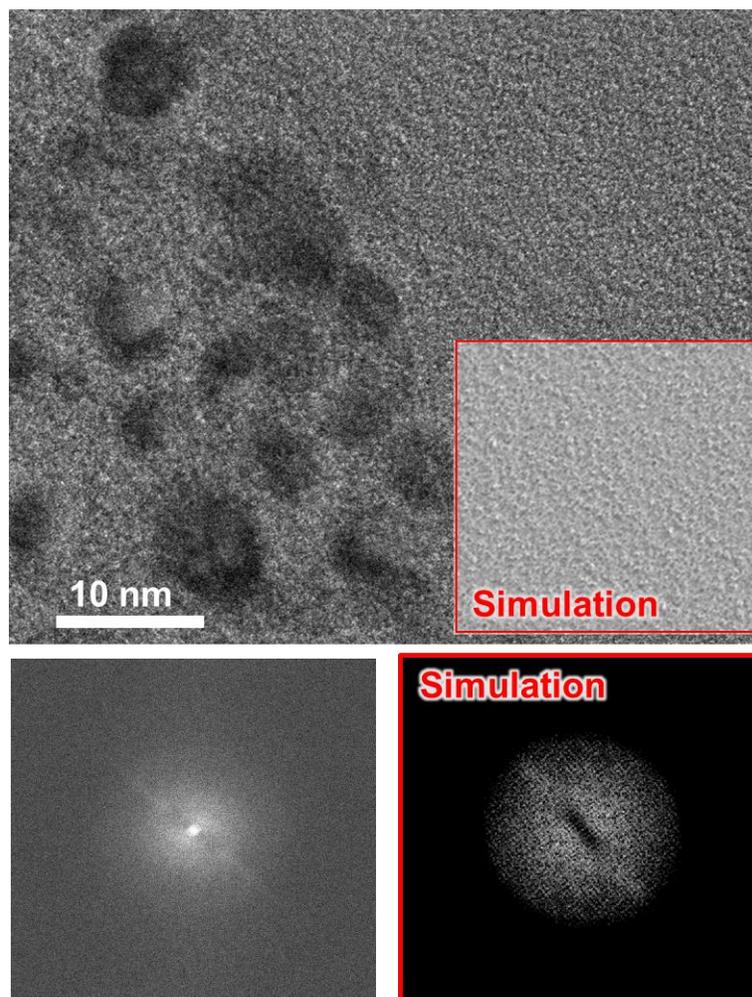

Figure S6. Direct comparison of the experimental and simulated images (upper row) and FT spectra (lower row).

**Probe shape**

Simulated and experimentally obtained probe images on the specimen plane are respectively listed in Figure S7 and S8. With a slight defocus semicircular aperture shape becomes apparent.

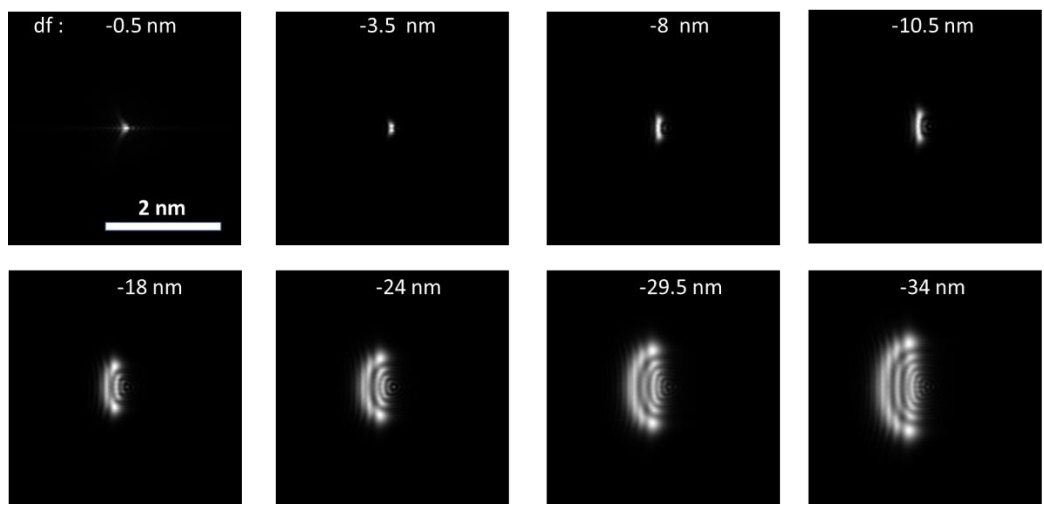

Figure S7. Simulated probe shape on the specimen plane using a semicircular aperture with different defocuses. Slight two-fold and three-fold astigmatisms are introduced to match the experimental results in Figure S8.

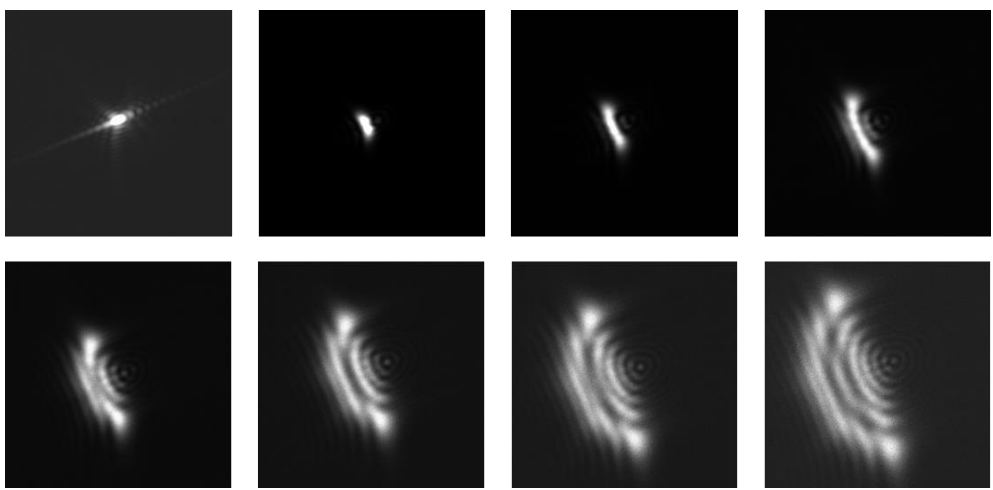

Figure S8. Experimentally obtained probe shapes on the specimen plane using a semicircular aperture with different defocuses.

**Line profile of the liver sample images**

The line profiles of the conventional and SCI STEM images are shown in Figure S9. The SCI-STEM profile is composed of both BF and differential signals.

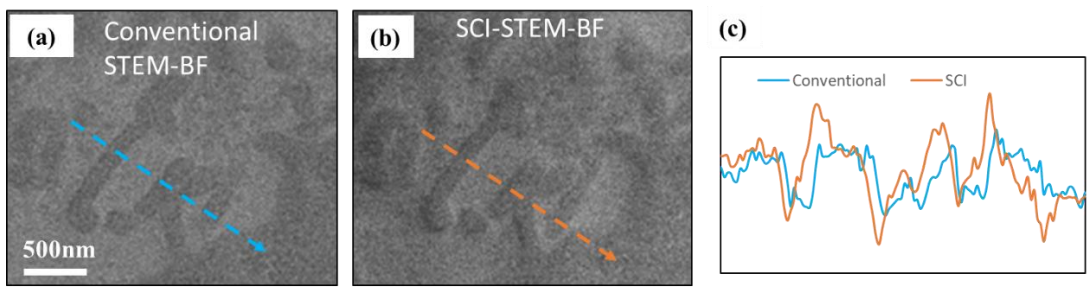

Figure S9. Line profiles of (a) conventional and (b) SCI-STEM images of the liver sample (same as Figure 6 in the main text). (c) Line profile along the "differential" direction. The contrast offset is adjusted for comparison. The profiles are integrated over the line width of 10 pixels.

**Detector size dependence**

As shown in Figure S3-5, with a larger detector, more conventional bright field (BF) component is included along the straight edge direction of the semicircular aperture. Figure S10 shows the detector-size dependence of this contrast component change, namely conventional BF and SCI components.

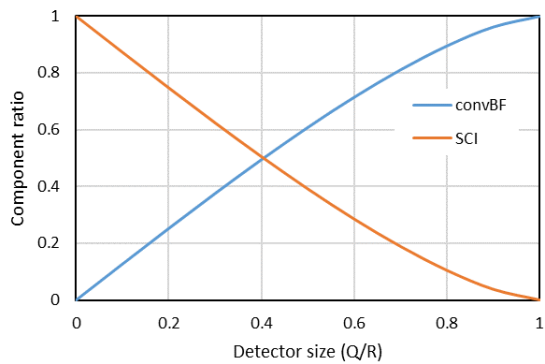

Figure S10. Component ratio as a function of the detector size. The detector size is normalized by the illumination aperture size.